\documentclass[conference]{IEEEtran}

\usepackage{cite}
\usepackage{amsmath,amssymb,amsfonts}
\usepackage{algorithm,algorithmic}
\usepackage{graphicx}
\usepackage{textcomp}
\usepackage{xcolor}
\usepackage{subfig}
\usepackage{array}
\usepackage{caption}

\usepackage{multicol}
\usepackage{enumitem}

\usepackage{amsmath}
\usepackage{mathtools}

\usepackage{graphicx}
\usepackage{epstopdf}
\usepackage{psfrag}

\usepackage{graphicx,xcolor} 

\usepackage{times}
\usepackage{url}

\usepackage{algorithm}
\usepackage{tabularx}
\usepackage{caption}
\usepackage{float} 
\usepackage{amssymb}
\usepackage{color}
\usepackage{multirow}

\usepackage{caption}
\usepackage{amsmath, amsthm, amssymb}
\usepackage{booktabs}
\usepackage{boxedminipage}
\usepackage{morefloats}
\usepackage{wasysym}

\begin{document}

\pagenumbering{arabic}

\title{Model and Machine Learning based  Caching and Routing Algorithms for Cache-enabled Networks}

\author{Adita Kulkarni, Anand Seetharam 
\\ Department of Computer Science, SUNY Binghamton, USA
\\{akulka17@binghamton.edu, aseethar@binghamton.edu}}

\maketitle 

%

\begin{abstract}
In-network caching is likely to become an integral part of  various networked systems (e.g., 5G networks, LPWAN and IoT systems) in the near future.  In this paper, we compare and contrast model-based and machine learning approaches for designing caching and routing strategies to improve cache network performance (e.g., delay, hit rate). We first outline the key principles used in the design of model-based strategies and discuss the analytical results and bounds obtained for these approaches. By conducting experiments on real-world traces and networks, we identify the interplay between content popularity skewness and request stream correlation as an important factor affecting cache performance.  With respect to routing, we show that the main factors impacting performance are alternate path routing and content search.  We then  discuss the applicability of multiple machine learning models, specifically reinforcement learning, deep learning, transfer learning and probabilistic graphical models for the caching and routing problem.
\end{abstract}


%
%


\section{Introduction}
\label{sec:intro}
Over the last decade, cache networking research  (e.g., information-centric networking) has gathered significant momentum and its  benefits are likely to impact a variety of future communication systems including 5G networks, clouds, LPWAN and IoT systems  \cite{paschos2018role, din2018caching}.  In fact,  information-centric architectures have already shown promising initial results in IoT systems \cite{pfender2018performance}.  By caching and serving content from in-network  nodes rather than the content custodians (origin servers),  cache-enabled networks   seek to improve  user performance. Therefore, a  number of  caching and routing strategies have been designed that effectively leverage in-network caching to improve performance.

Therefore, in this  paper, we outline the key principles used in the design of these protocols and quantitatively demonstrate how these principles aid in improving performance. Based on prior work, we identify three main approaches for developing caching and routing protocols---{\it i)} Design optimized cache management strategies assuming that requests for content are routed according to the network's underlying routing strategy. {\it ii)} Design optimized routing strategies assuming that the network adopts some native cache management strategy. {\it iii)} Design strategies that jointly optimize for caching and routing.  We then present research on the analysis of caching and routing protocols that complement and aid the understanding of the  design factors required for developing new protocols. In particular, we present an overview of recent  analytical research that seek to answer questions related to optimality,  performance guarantees and attempt to determine the actual performance of particular protocols in  specific settings. 

We  conduct experiments using multiple real-world networks and traces and show that the interplay between content popularity skewness and request stream correlation is an important factor affecting cache performance.  We also demonstrate that augmenting shortest path routing with alternate path routing and content search can significantly improve performance.




We next present  an overview of  machine learning approaches that have been used to address the caching and routing problem in cache-enabled networked systems. We discuss the potential benefits of multiple different classes of machine learning algorithms, in particular reinforcement learning, deep learning, deep reinforcement learning, transfer learning,  and probabilistic graphical models to solve the  caching and routing problem. We conclude by discussing the various challenges that need to be overcome to allow the seamless adoption of machine learning models to solve these problems.

{\it The goal of this  paper is to provide an overview of state-of-the-art research in cache networks in a succinct manner, to draw attention to  key contributions in the field, to highlight the various model-based and machine learning approaches that can be used to solve the caching and routing problem and to stimulate further discussion.} 


\section{Key Design Principles of Routing and Caching}
\label{sec:background}
In this section, we provide an overview of the main design considerations while  developing new caching and routing protocols. We first outline the principles behind designing caching protocols, followed by routing and conclude by discussing joint caching and routing. Table \ref{tab:route-cache} provides an overview of some of the recently proposed caching and routing algorithms. Due to lack of space, we are unable to cite each paper individually. Most citations can be found within \cite{paschos2018role}.

\subsection{Caching}

When adopting a cache management strategy, the network has two options---static caching/content placement or dynamic caching. We first describe static and dynamic caching and then highlight the differences between them. 

\subsubsection{Static Caching}

If static caching is adopted, the network decides the set of content to be placed at the different network nodes so as to optimize performance. The set of content to be placed at the various network nodes is determined a priori,  primarily based on the content popularity and then these pieces of content are  cached at network nodes. As network caches do not change their cached content, in static caching, requests for cached content result in hits  while requests for all other  content result in misses.  While for the single cache case, the optimal static caching strategy is caching the most popular content, for an interconnected network comprising of multiple nodes, determining the set of content to cache, particularly at core network nodes is considerably harder. As upstream caches only receive the miss request stream from downstream caches, this miss stream dictates the content placement at these nodes. Though determining the optimal set of content to cache in a general network is still largely unsolved, Banerjee et al. propose a greedy solution to this problem \cite{banerjee2018greedy}. 



\subsubsection{Dynamic Caching}

In dynamic caching, the content of network caches can potentially change as new content passes through them. Dynamic caching strategies thus have two important decisions to make, one is cache eviction and  the other is cache insertion. 

\noindent {\it Cache Eviction:} If a node decides to cache a particular content, the cache eviction policy decides what content to evict from the cache. Popular cache eviction policies are the Least Recently Used (LRU) and First In First Out (FIFO) policies.

\noindent {\it Cache Insertion:} The other important aspect of dynamic caching is cache insertion that aims to improve performance by increasing the network content diversity as well as by pushing popular content closer to the user. The simple Leave Copy Everywhere (LCE) policy results in a piece of content being cached at all nodes on the return path from the custodian. To increase  network content diversity, two widely adopted metrics are---{\it i)} network centrality that uses the centrality of nodes to determine what content to cache, and {\it ii)} a probabilistic approach that takes factors such as content popularity, node  connectivity and whether other nodes on the path have cached the same content  into account to determine if a content should be inserted into a cache. Cache Less for More (CL4M) and ProbCache are two popular strategies that rely on  network centrality and adopt a probabilistic approach, respectively.

\begin{table*}[!ht]
  \centering
  \small{
  \begin{tabular}{| l   r | c | c | c |}
  	\hline
	 \multicolumn{2}{| c |}{Protocol Name} & {\parbox{1.5cm}{\centering Type}} & {\parbox{ 2.5 cm}{\centering Machine Learning}} &  {\parbox{ 4.5 cm}{\centering Summary of Contributions}} \\ 
	 & & & based Approach & \\
	 \hline
	 \multicolumn{2}{| l |}{\textit{Greedy Caching \cite{banerjee2017greedy}}}  & Static Caching & No & Exploits content popularity   and miss stream from downstream \\
	 & & & & nodes to make caching decisions in a general network. \\
	 \hline
	 \multicolumn{2}{| l |}{\textit{Femto Caching}}  & Static Caching & No  & Exploits content popularity to make  caching decisions  in a \\
	  & & & &  heterogeneous cellular network with performance guarantees. \\
	 \hline
	  \multicolumn{2}{| l |}{\textit{Least Recently}}  & Dynamic Caching  & No & Evicts the content that has not been accessed   \\
	  {\textit{Used (LRU)}}  & &  (Cache Eviction) &  & for the longest time duration. \\
	 \hline
	 \multicolumn{2}{| l |}{\textit{Leave Copy}}  & Dynamic Caching & No &  Cache a copy of the content on all en route caches.\\
	  {\textit{Everywhere (LCE)}}  & &  (Cache Insertion) &  &\\
	 \hline
	 \multicolumn{2}{| l |}{\textit{Leave Copy}}  & Dynamic Caching  &  No &  Cache a copy of the content at the node  that is one \\
	 {\textit{Down (LCD)}} & &  (Cache Insertion) & & hop downstream from the cache hit.\\
	 \hline
	 \multicolumn{2}{| l |}{\textit{Cache Less for}}  & Dynamic Caching &  No & Cache content based on network centrality. \\
	{\textit{More (CL4M)} }  & &  (Cache Insertion) & & \\
	 \hline
	 \multicolumn{2}{| l |}{\textit{PopCache}}  & Dynamic Caching &  No & Cache content based on popularity. \\
	 & &  (Cache Insertion) & &\\
	 \hline
	 \multicolumn{2}{| l |}{\textit{ProbCache}}  &  Dynamic Caching &  No & Probabilistically cache content at a node \\
	 & &  (Cache Insertion) & & \\
	 \hline
	  \multicolumn{2}{| l |}{\textit{Hybrid Caching\cite{kulkarni2018hybird}}}  &  Combines Static and  &  No &  Divide caches into static and dynamic components\\
	  & &  Dynamic Caching & & based on a utility function. \\
	 \hline
	  \multicolumn{2}{| l |}{\textit{Hash-Routing}}  &  Routing &  No & Route requests based on hash tables. \\
	 \hline
	  \multicolumn{2}{| l |}{\textit{Breadcrumbs} }  &  Routing &  No & Uses pointers (breadcrumbs) to keep track of content, \\
	  & &  & & follow pointers to obtain content.\\
	 \hline
	 \multicolumn{2}{| l |}{\textit{CTR} \cite{banerjee2017characteristic}}  &  Routing & No &  Uses characteristic time of a content in a cache to route requests.\\
	 \hline
	 \multicolumn{2}{| l |}{\textit{Optimal Caching} \cite{dehghan2017complexity}}  &  Joint Routing & No & Determine the optimal set of content to be cached  \\
	 & &  and  Caching &  & and the routing strategy  adopted by each node  \\
	    & &   & &  by taking network congestion into account. \\
	 \hline
	 \multicolumn{2}{| l |}{\textit{DeepCache}\cite{narayanan2018deepcache}}  &  Dynamic Caching &  Yes  &  Deep LSTM based encoder-decoder model \\
	 & &   & & to predict the request stream,\\
	 & &   & & smart caching policy based on these predictions.\\
	 \hline
	 \multicolumn{2}{| l |}{\textit{Q-Caching}}  &  Dynamic Caching &  Yes &  Uses Q-learning based approach\\
	 & &   & & to determine what content to cache.\\
	 \hline
  \end{tabular}
  \caption{A Comparison of Caching and Routing Algorithms}
  \vspace{-0.5 cm}
  \label{tab:route-cache}
  }

\end{table*} 

\noindent {\it Joint Cache Insertion and Eviction:}  A variety of policies have also been proposed  that  address the cache insertion and eviction aspects together. For example, a number of different variants of LRU (e.g., $p$-LRU, $k$-LRU \cite{garetto2016unified}) have also been proposed that address the cache insertion aspect assuming that the eviction policy is LRU. In $p$-LRU, a piece of content is inserted  into a cache based on some probability $p$, while the  $k$-LRU policy exploits a chain of $(k-1)$ virtual caches to filter content. Before a request arrives at the physical cache that stores the  actual content, it passes through a chain of $(k-1)$ virtual caches that are in front of it. These virtual caches only store object pointers and perform caching operations on them. A content or a pointer can only be stored in the cache at level $i$ if it obtains a hit at level $(i-1)$.

\subsubsection{Static vs. Dynamic Caching}

A natural question that arises is what are the advantages of adopting one type of caching strategy (i.e., static or dynamic) over the other? To answer this question, it is important to understand how static caching and dynamic caching attempt to serve requests.  Static caching strategies take advantage of the long tail of the content popularity distribution (i.e., small number of content receive majority of the requests) and cache popular content within the network, while dynamic caching leverages the correlation in the request stream.  Therefore, the performance of static and dynamic caching strategies is dominated by the interplay of the skewness of the popularity distribution and the request stream correlation. Another important question that arises is how to adapt static caching to real-world scenarios where content popularity varies over time? In such scenarios, the  approach adopted by static caching is to estimate content popularity over a certain time window and to cache content based on it. This process is repeated periodically to help static caching capture temporal variations in popularity. 





\subsection{Routing}
\label{sec:route}



Having studied the main principles adopted for designing caching strategies, in this section, we focus on routing.  A key idea to effectively utilize in-network caches is to seek alternate paths for obtaining content in addition to the shortest path. In this context, the simplest approach is to adopt standard multi-path routing. A smarter approach is to perform bookkeeping in the form of keeping breadcrumbs (i.e., pointers) at users and intermediate routers in order to keep track of the node(s) from where a particular content is recently obtained. By following the trail of these breadcrumbs, a node can potentially obtain  content faster than shortest path routing. Content search,  particularly in mobile networks is another key principle that is used in conjunction with shortest path routing to improve performance \cite{banerjee2018content}. It exploits the fact that  the requested content may be cached nearby and thus readily available at neighboring nodes. 

\subsection{Joint Caching and Routing}

Instead of focusing solely on caching or routing, recent research has also tried to jointly optimize caching and routing  \cite{dehghan2017complexity}. While solving the joint problem, majority of existing approaches attempt to find the optimal content placement and routing  and do not approach the problem from the dynamic caching perspective. Based on  previous research, we next outline the basic steps generally adopted by the research community to solve the joint caching and routing problem. 


\begin{itemize}[leftmargin=*]
\item The usual methodology adopted is to cast the joint content placement and routing problem as an optimization one subject to constraints such as  the caching capacity at various nodes and the connectivity among the different nodes. 
\item The main objective functions considered in prior work are delay and hit rate with some recent research also focusing on general utility functions \cite{dehghan2017complexity}. 
\item Prior research has also demonstrated that most of these formulations turn out to be computationally hard (i.e., NP-hard) \cite{dehghan2017complexity}.  One of the main factors contributing to the hardness is the fact that finding the optimal content placement results in a combinatorial explosion. 
\item A natural next step is to formulate approximation algorithms that are computationally efficient and provide performance guarantees. Most of these problem formulations are integer linear programs, thus making them amiable to approximation algorithms. One approach is to demonstrate that the objective function is submodular and that the constraints follow a matroid. This subsequently entails that there exists a greedy solution that provides a $(1-1/e)$ approximation guarantee. Additionally, researchers have designed heuristic solutions that provide good performance in practice.
\item The proposed approximation and heuristic solutions are generally centralized in nature which necessitates that the problem be solved in a central server and the results be distributed to network nodes. To address this concern, several efficient distributed solutions have also been proposed recently. A widely adopted technique is to design a gradient descent/ascent approach that asymptotically converges to the same solution as obtained by the centralized approach.
\end{itemize}

\subsection{Analysis of Caching and Routing}


While understanding the key factors  governing performance is necessary to develop novel caching and routing strategies, analysis is essential to quantify the performance of algorithms in particular network settings and understanding the scenarios where one strategy is likely to outperform another. Additionally, analytical bounds and expressions can also be used to design better caching and routing strategies and can aid network management and maintenance. 


\begin{table*}[!ht]
  \centering
  \small{
  \begin{tabular}{| l   r | c |}
  	\hline
	 \multicolumn{2}{| c |}{Approach}  &  {\parbox{ 4 cm}{\centering Summary of Contributions}} \\ 
	 \hline
	 \multicolumn{2}{| l |}{\textit{Che et al.'s Approximation} } & Determines the hit rate of LRU caches. \\
	 \hline
	  \multicolumn{2}{| l |}{\textit{Rizk et al.'s Approximation \cite{rizk2017model}}}  & Determines the hit rate at LRU caches in cache hierarchies described by a  directed acyclic graph.   \\
	 \hline
	 \multicolumn{2}{| l |}{\textit{a-Net} \cite{rosensweig2012analysis}}  &  Iterative algorithm to determine the hit rate of a network of LRU caches. \\
	 \hline
	  \multicolumn{2}{| l |}{\textit{Garretto et al. 's Approximation} \cite{garetto2016unified}}  &  Extends Che's approximation to determine the hit rate  of FIFO and RANDOM cache\\
	   & &  eviction  policies. Also determines expressions for LCE, LCD and LCP cache insertion policies. \\
	 \hline
	  \multicolumn{2}{| l |}{\textit{Approximation of TTL caches}}  &  Determine how to set the parameters of TTL caches to mimic the behavior of other policies\\
	  & & such as FIFO and LRU. \\
	 \hline
  \end{tabular}
  \caption{Analytical Approaches for Determining Cache Hit Rate}
  \vspace{-0.5 cm}
  \label{tab:hit-rate}
  }
\end{table*}

\subsubsection{Caching} A significant amount of effort has been invested in understanding the performance, in particular the network hit rate for different cache insertion and eviction policies. Table \ref{tab:hit-rate} succinctly describes the research in determining the hit rate of network caches. In one of the seminal papers, Che et al. derive  approximations for the hit rate of LRU caches. This approximation, popularly known as Che's approximation has been shown to be applicable for general content popularity distributions. In recent years, this approximation has been extended to non-stationary requests and to general networks comprising of multiple nodes.  Garretto et al. \cite{garetto2016unified}  derive expressions for the hit rate for  multiple caching insertion and eviction policies such as LRU, $p$-LRU, $k$-LRU, FIFO, LFU and RANDOM, and LCE, Leave Copy Down (LCD) and Leave Copy Probabilistically (LCP) respectively. Simulation and trace-based evaluation show that the analytical and  simulation results match closely. This study also demonstrates the superiority of the $k$-LRU policy in comparison to other strategies. Alongside, research effort has also been devoted to analyze the performance of Time To Live (TTL) based caches because in general it is easier to derive exact expressions for uncorrelated and correlated request streams. 

In \cite{qiu2016cache, qiu2017popularity} the authors  analytically study the fundamental limits of caching in wireless networks. For example, the authors in \cite{qiu2016cache} obtain upper bounds on capacity  and achievable capacity lower bounds in wireless cache networks. Similarly,  in  \cite{qiu2017popularity}, the authors investigate the capacity  scaling laws in cache-enabled wireless networks considering the skewness of the Zipfian popularity distribution. They show that the capacity at individual nodes increases monotonically with the number of nodes for skewed popularity distributions.  These scaling laws are invaluable and help us appreciate the maximum benefits of caching.  Similarly, the authors investigate a general cache network using queuing theory and determine how to place objects in caches to attain a desired objective \cite{kelly2019}.




\subsubsection{Routing} Theoretical analysis has also been conducted to determine the extent to which content search and scoped flooding is beneficial. Analysis and experiments show that the optimal flooding radius is small (less than 3 hops). This means that flooding requests beyond the immediate neighborhood of a requester is likely to incur significant overhead while providing minimal performance improvement. The benefits of opportunistic routing, an important routing paradigm designed for wireless networks that exploits the broadcast nature of the wireless medium to select the best relay to forward a request toward the custodian has been analyzed in \cite{herath2018analyzing}. The authors design Markovian models to analyze the performance of opportunistic request routing in wireless cache networks in the presence of multi-path fading. Based on their results, the authors conclude that the benefits of in-network caching are more pronounced when the probability of successful packet transmission is low. This result  suggests that caching is likely to have more benefits in a wireless network with lossy links. 

Popular implementation of the ICN architecture such as Content-centric Networks (CCN) and  Named Data Networks (NDN) perform aggregation of requests for same content (popularly known as Interest aggregation) through the use of a data structure called Pending Interest Table (PIT) to improve routing performance. A recent study investigates and quantifies the benefits that PITs provide under realistic conditions. This preliminary investigation suggests that only a small fraction of requests may benefit from request aggregation with the benefits being closely tied to the network cache budget. 




\section{Experimental Results}

In this section, we experimentally demonstrate the performance benefits of the key principles discussed  in the previous section in the design of caching and routing strategies. To demonstrate how the interplay of content popularity skewness and correlation impact the performance of caching strategies, we conduct experiments on multiple real-world topologies (e.g., GARR, WIDE, GEANT), synthetic and real-world request stream traces (e.g., YouTube, Wikipedia) and multiple cache insertion policies (LCE, CL4M and ProbCache). We present representative results for the GARR topology and the YouTube trace to avoid cluttering the paper with multiple similar figures.   The GARR topology comprises of 61 nodes and 21 users. The YouTube request stream trace used here was collected over a campus network at the University of Massachusetts Amherst. In this particular trace, the long-term content popularity is low whereas the overall correlation among requests is high, which means that  requests for the same content tend to occur in bursts. We assume  all content to be of unit size and vary the cache size in our experiments.

\begin{figure}
    \centering
  \subfloat[Synthetic Request Stream]{%
       \includegraphics[scale=0.26]{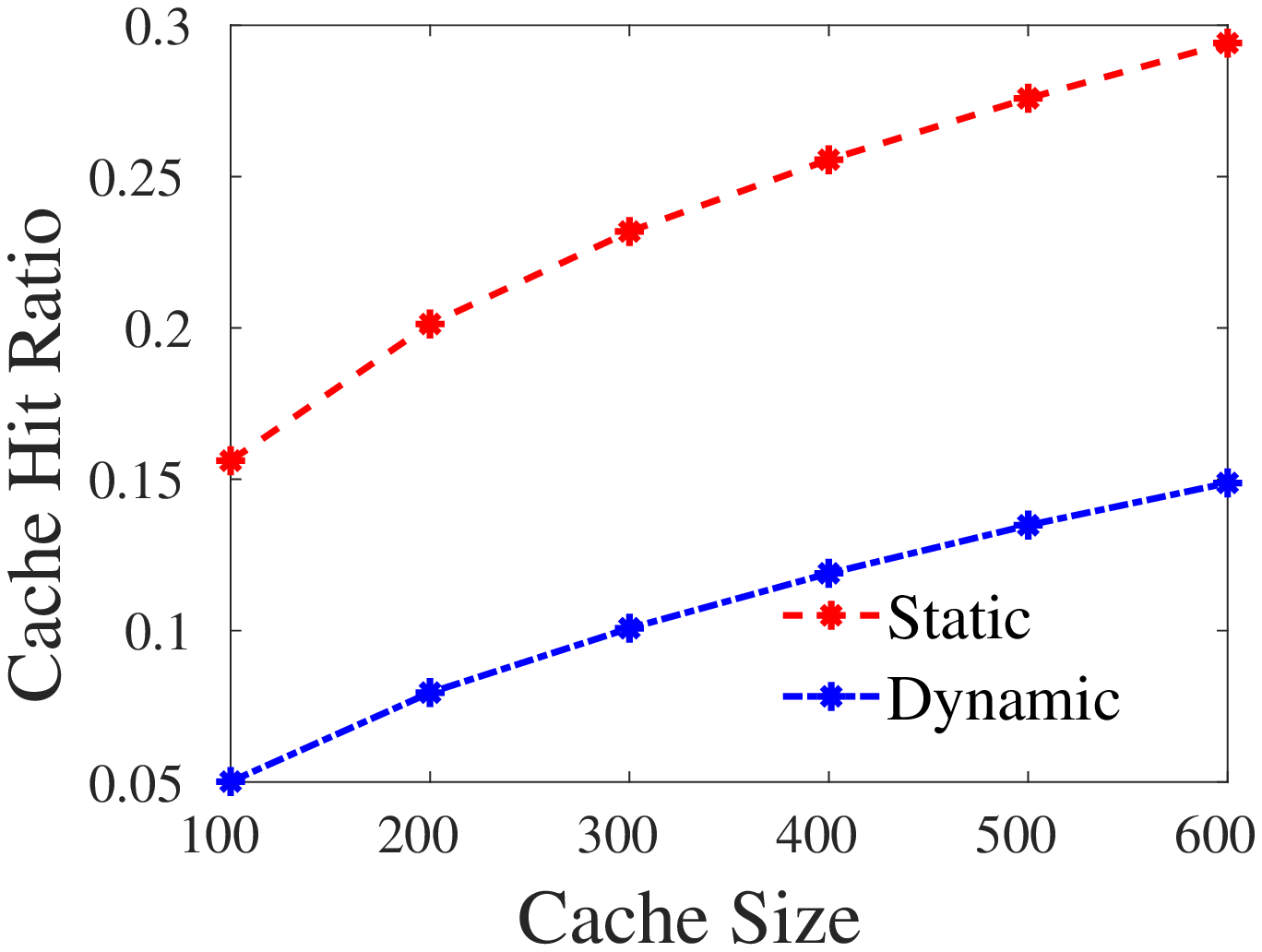}
    \label{fig-synthetic}}    
    \hspace{2mm}
  \subfloat[YouTube Request Stream]{%
       \includegraphics[scale=0.26]{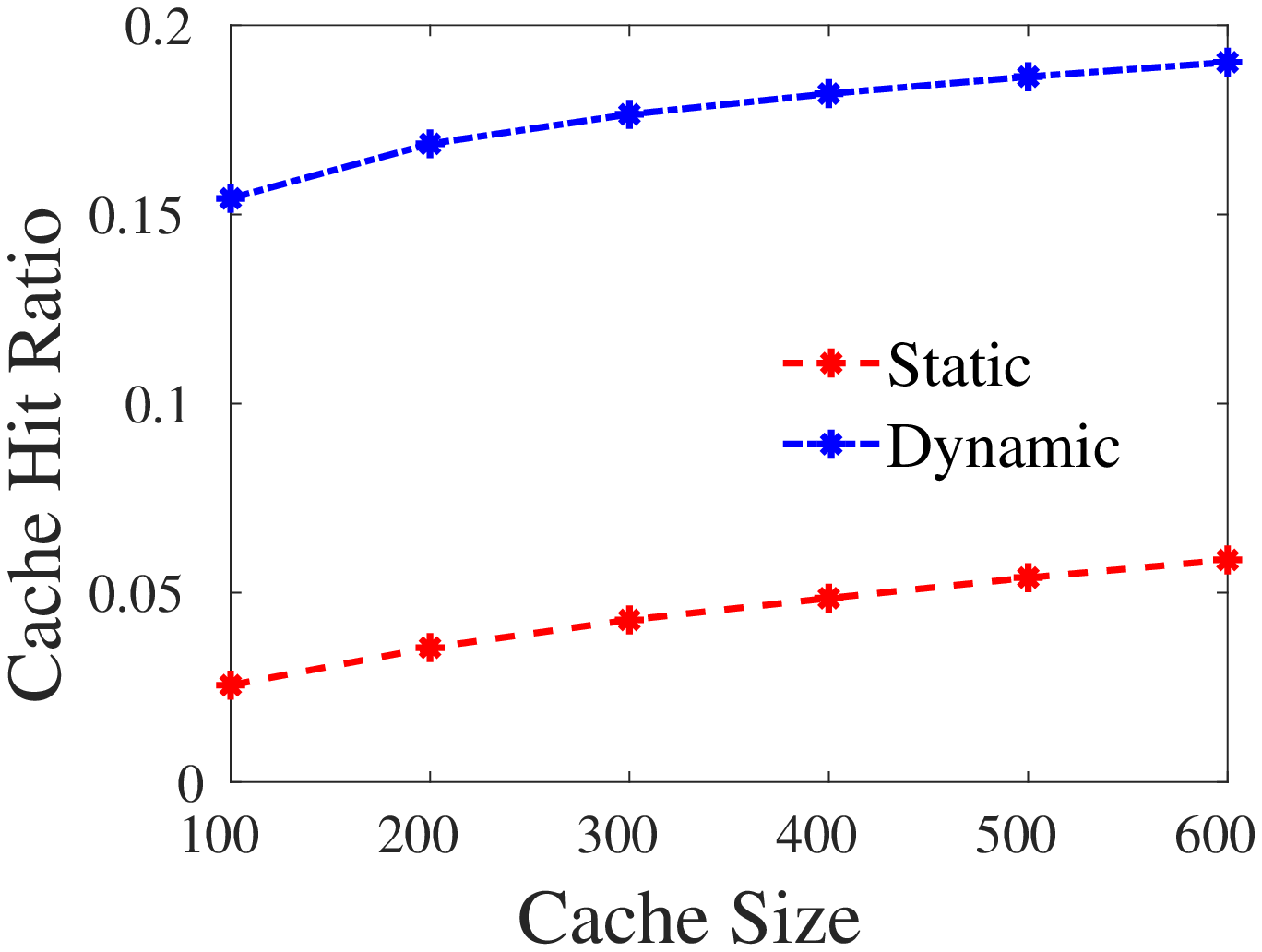}
    \label{fig-realworld}
}
  \caption{Performance of Static and Dynamic Caching}
  \vspace{-3mm}
  \label{fig-synthetic-real} 
\end{figure}

\begin{figure}
    \centering
  \subfloat[Network Content Diversity]{%
       \includegraphics[scale=0.26]{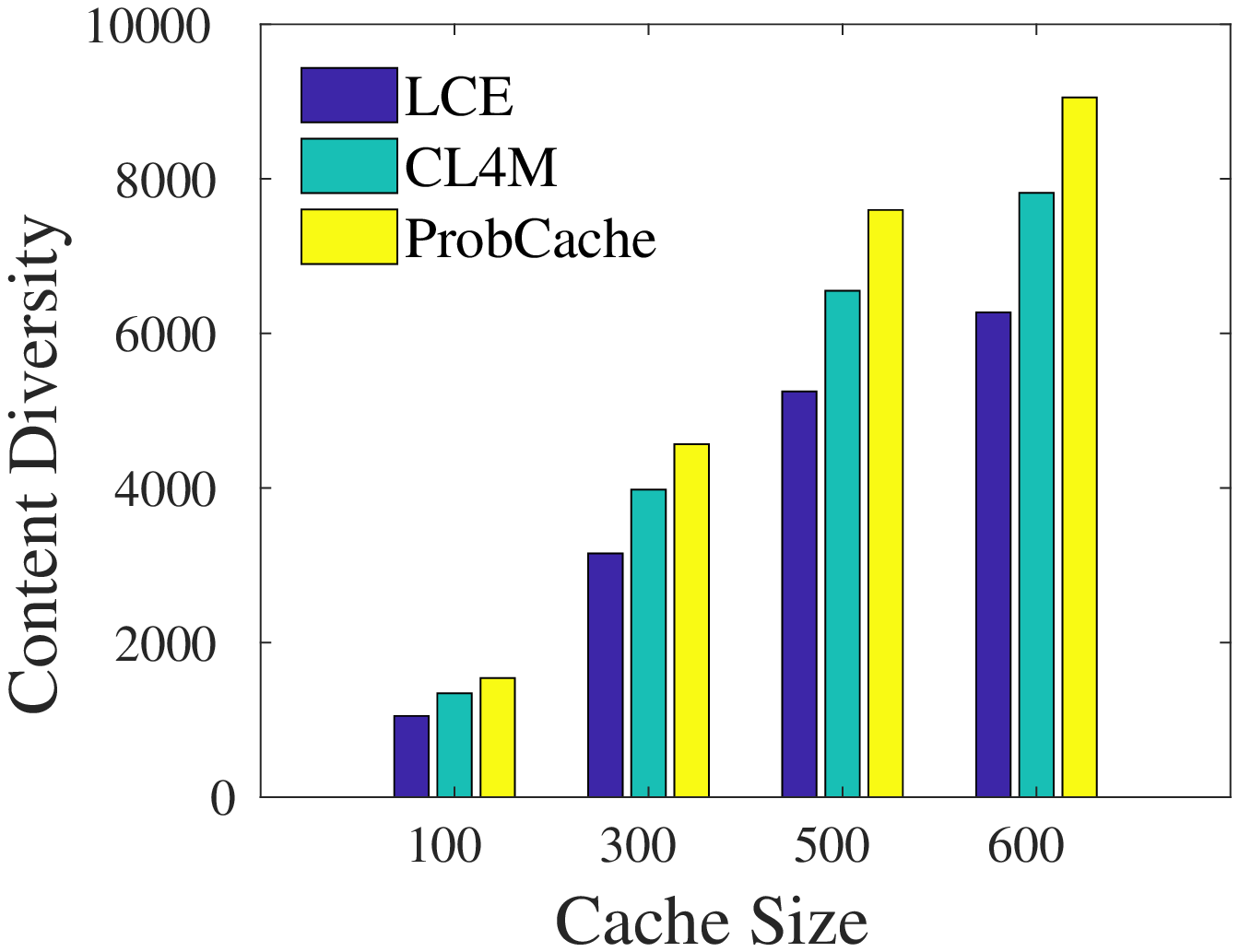}
    \label{fig-content-diversity}}    
    \hspace{2mm}
  \subfloat[Performance Impact of Multi-path Routing and Content Search]{%
       \includegraphics[scale=0.26]{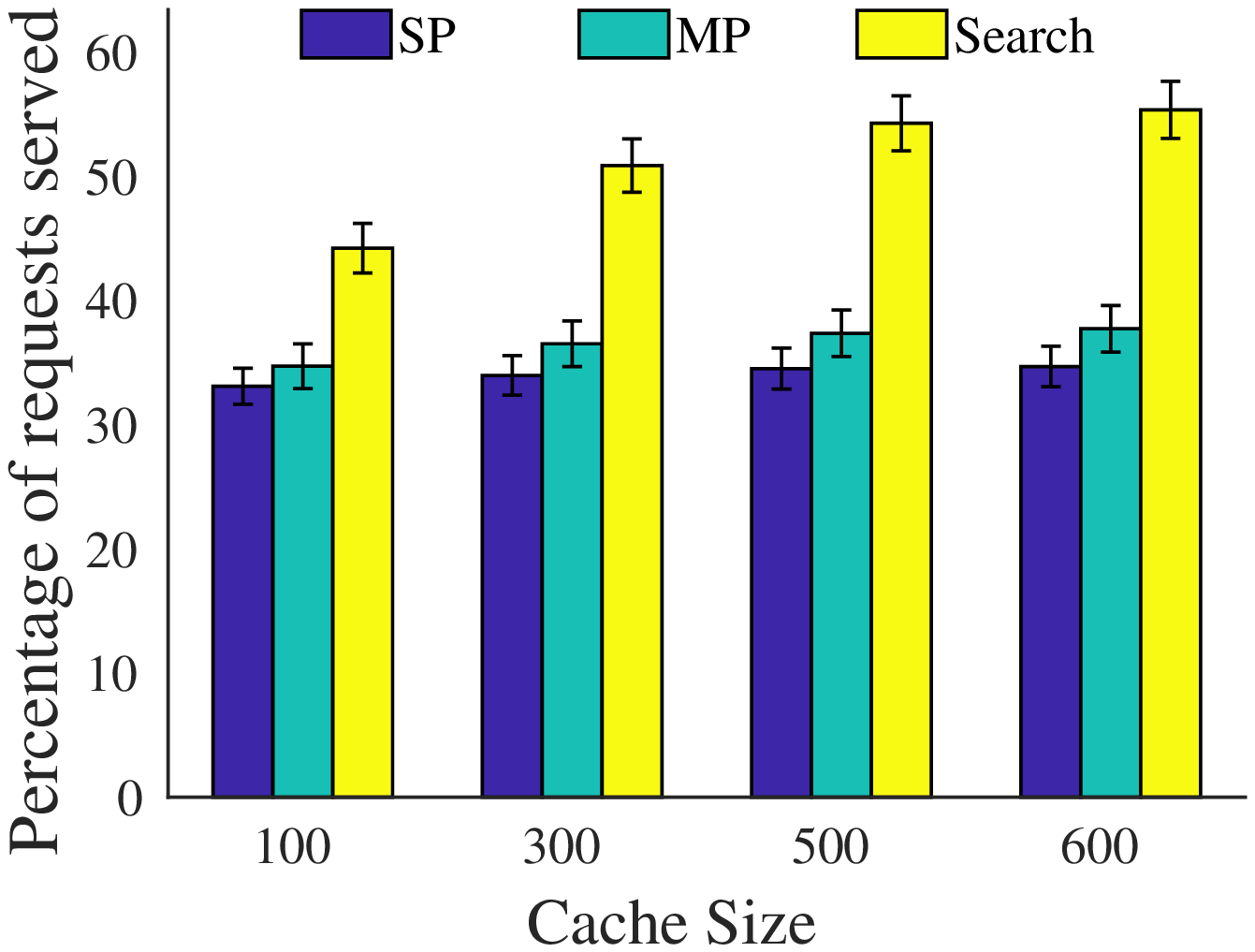}
    \label{fig-search}
}
  \caption{Comparing  Caching and Routing Strategy Performance}
  \vspace{-3mm}
  \label{fig-routing} 
\end{figure}

To study the impact of request stream  on the performance of static and dynamic caching, let us consider Figures \ref{fig-synthetic} and \ref{fig-realworld}.  In Figure \ref{fig-synthetic}, we assume that content popularity is distributed according to a Zipfian distribution with skewness parameter $\alpha$ = 0.7 and requests for content are generated following an independent request stream model, while Figure \ref{fig-realworld} is generated for the YouTube trace. The static caching policy adopted in the figures is Greedy Caching  \cite{banerjee2018greedy}, while the dynamic cache insertion and eviction policies are Leave Copy Everywhere (LCE) and LRU respectively. We observe from Figure \ref{fig-synthetic} that static caching outperforms dynamic caching while the opposite is true in Figure \ref{fig-realworld}. This is because as Figure \ref{fig-synthetic} is generated considering an identically and independently distributed request stream, the correlation is 0 and hence, static caching outperforms dynamic caching. In comparison, as the overall correlation in the request stream increases, as is the case in the YouTube trace, dynamic caching outperforms static caching (Figure \ref{fig-realworld}).

%
%
%


We next turn our attention to dynamic caching policies and study the impact of cache insertion policies on content diversity, which in turn affects performance.  We define content diversity as the   total number of unique content in the network. Figure \ref{fig-content-diversity} shows the content diversity for LCE, CL4M and ProbCache for various cache sizes for the GARR topology for the YouTube trace. We observe from Figure \ref{fig-content-diversity} that the content diversity for CL4M and ProbCache is considerably higher in comparison to LCE.  This content diversity also translates to improved performance (e.g., hit rate), with both ProbCache and CL4M  outperforming LCE \cite{banerjee2018greedy}. 

From the above discussion, it is evident that static and dynamic caching attempt to exploit different aspects of the request stream to improve performance. Recent work has  explored the benefits of hybrid caching that combines the best aspects of static and dynamic caching \cite{kulkarni2018hybird}. The key idea is to split a cache into two parts---a static part that statically caches content based on popularity and a dynamic part that leverages content popularity.  Determining the optimal split is an important research question that is still being investigated.

We next study   the positive performance impact of ideas such as multi-path routing  and content search (Figure \ref{fig-search}). As paths to content custodians are always available in a static network, we consider a real-world mobile network comprising of pedestrian users to demonstrate their benefits. To this end, we consider the  Stockholm pedestrian mobility trace that   contains simulation traces of pedestrians walking in a part of downtown Stockholm, covering an area of 5872 sq. m. For our experiments, we consider 300,000 location entries consisting of 587 pedestrians.  As mobility can result in frequent path breakages, Figure \ref{fig-search} shows the percentage of requests served for various cache sizes if we adopt either  multi-path or content search to execute on top of shortest path routing. We observe from the figure that both  multi-path (denoted by MP) and content search aid in serving a greater number of requests. We observe that using shortest path (denoted by SP) and content search together serves significantly more number of requests than multi-path forwarding. The reason for the limited performance improvement in multi-path routing is that both paths in multi-path routing are calculated based on the prior network state and thus when the shortest path breaks due to node mobility, the alternate path to the custodian is also likely to be broken. In comparison, large number of requests can be served by  content search  because node mobility helps in exploiting content diversity by searching new neighbors.


\section{Machine Learning Approaches for Caching and Routing}

In this paper, so far we have focused on design principles and analysis of caching and routing strategies. In this section, we first investigate the benefits of adopting machine learning techniques to solve the caching and routing problem and then list some of the challenges that need to be overcome to allow the seamless adoption of machine learning for these problems. 

\subsection{Possible Machine Learning Models}
The performance of caching strategies can be improved if one can accurately predict  changes in content popularity and determine what content is likely to be requested in future. Learning web request streams and using them to improve the performance of HTTP caches   has been well investigated in the early part of this century. Models such as n-gram models, Markov models and Markov trees  developed in the context of HTTP caches can also be adopted in cache networks with certain modifications.

\noindent {\it Reinforcement Learning:} The  necessity to make joint caching and routing decisions  to improve performance makes this problem an ideal candidate for adopting reinforcement learning methods. To this end, a reinforcement learning method called Q-caching  has been proposed that builds on standard Q-routing to make joint caching and routing decisions.   Q-caching increases network content diversity, reduces the load at custodians and content download times for clients.   Similarly, in a recent work, the authors propose a reinforcement learning approach that uses model-free acceleration for online cache replacement by taking predicted content popularity,  cache hits and replacement costs  into account.

\noindent {\it Deep Learning:} Deep learning models have also been proposed to predict future content popularity variations by taking the spatial-temporal features of popularity into account. For example, a recently proposed approach DeepCache \cite{narayanan2018deepcache} uses a deep LSTM based encoder-decoder model  to predict the changes in content popularity. The authors then design a caching policy, which takes predicted content  popularity information into account to make smart caching decisions. Similarly, in other recent work, lightweight cache insertion and eviction schemes that take these deep learning predictions in consideration have also been proposed.

\noindent {\it Deep Reinforcement Learning:} In recent years, multiple deep reinforcement learning approaches  (e.g., asynchronous advantage actor-critic (A3C), deep deterministic policy gradient (DDPG), Deep Q Networks (DQN), Trust Region Policy Optimization (TRPO))  have been designed that have been shown to provide good performance in multiple different domains.   Most of the above-mentioned approaches are based on deep actor-critic architectures,  a recent architectural improvement to reinforcement learning. Actor-critic architectures are also the easiest to adapt to an online training setting. For example, the asynchronous nature of A3C helps in designing parallel and distributed implementations of the algorithm that allow for greater exploration of the state space, resulting in good test performance on previously unseen data. Similarly, the presence of off-policy updates in DDPG similarly allow for a wide exploration of the state space. DDPG also has provisions to learn from a large amount of past data and uncorrelated transitions from the replay buffer.  


\noindent {\it Transfer Learning and Bandit Models:} Along with reinforcement learning, transfer learning and bandit models are also good candidates for determining changes in content popularity and making joint caching and routing decisions.  The main idea behind  transfer learning is to leverage the knowledge gained in one domain and apply it to related but `new' domain. In this regard, content popularity estimation and in turn caching performance can potentially be improved by taking into account information such as a user's location and  social networking connections. To apply bandit models to cache networks,  we can assume the agent to have only  partial knowledge of the network. Based on this current knowledge, the agent takes actions to maximize its accumulated reward while acquiring new knowledge.

\noindent {\it Probabilistic Graphical Models:}  Probabilistic graphical models (both  discriminative and generative)  are  also good candidates for predicting content popularity variations \cite{koller2009probabilistic}.  Moreover, the low computational requirements of graphical models during the training phase also make them a more attractive option than deep learning models. While discriminative models (e.g., conditional random fields (CRFs)) only learn the conditional dependencies in the output variables (future predictions) given the input variables (past data) at training time, generative models (e.g., Markov random fields (MRFs), hidden Markov models (HMMs)) learn the joint dependencies in the entire data.  While at a cursory glance, generative models may appear to be superior than discriminative models as they jointly model the dependencies in the entire data, prior work has demonstrated that discriminative models often achieve superior prediction performance as they are tuned to maximize performance by learning structured outputs. In comparison, generative models capture the inherent dependencies in the data, help in accurately generating the data, and thus enable us to better understand the underlying network characteristics. 

\subsection{Challenges}

One of the biggest hurdles in adopting machine learning models, in particular deep models is the high computational resource requirement necessary for training these models. This computational overhead makes it harder to train the models in an online manner.  Apart from the computational overhead, another well-known issue with deep learning models is their lack of interpretability. Due to the complex inter-connection of cells in a neural network architecture as well as the large number of hidden layers, it is often very difficult to explain the
predictive performance of deep learning models. Another important issue that has been largely overlooked in prior research is the applicability of a trained model for a variety of different settings. In majority of prior work, even though the models are trained and tested on different data, both the train and test data are usually collected in a similar setting.  In our preliminary investigation, we have found that a  model trained in one setting may not necessarily perform well in a new setting. The main reason is that the sequences and data seen by the model at test time may not be similar to the ones seen at training time, thus leading to poor performance at test time. A fundamental question that arises in this regard is---how to design training datasets so that one can obtain overall good performance even in previously unseen network settings?

\label{sec:ml}

\section{Conclusion}
In this paper, we highlighted the key principles adopted in the design of  model-based caching and routing strategies to improve performance in cache networks. We then discussed the applicability of machine learning models, in particular deep learning, reinforcement learning, transfer learning and probabilistic graphical models for this problem.

\label{sec:conclusion}

\bibliographystyle{IEEEtran}
\bibliography{paper}

\end{document}